\documentclass[aps,prd,twocolumn,showkeys,floats]{revtex4}
\usepackage{graphics,epsfig,color} 
\usepackage{amsmath,amssymb}

\begin{document}
\title{A 2 PN/RM metric of General Relativity}
\author{Olivier Minazzoli\footnote{also in the IRAP PhD program} and Bertrand Chauvineau}
\affiliation{UNS, OCA-ARTEMIS UMR 6162}
\begin{abstract}
We present a derivation of 2PN/RM metric field equations from the Einstein field equation in General Relativity. We use the exponential parametrization and the $c^{-3}$ isotropic spatial coordinates such as in IAU2000 recommendations.
\end{abstract}
\keywords{post newtonian relativistic motion; post newtonian; harmonic gauge; TIPO; ASTROD; laser telemetry in the solar system}
\maketitle


\section{Introduction}
In a foreseeable future, telemetry will reach a new accuracy using laser links and high accuracy clocks (such as the TIPO experiments which is in study \cite{TIPO}). At such experiments accuracies, the effective 1,5 PN/RM (acronym defined in section \ref{sec:term}) metric defined from the 1PN IAU2000 \cite{IAU2000} metric will be no longer enough to fit data (2 PN/RM effects lead to terms of about $10^{-10}$s for a flying time of about 1000s when tagging units accuracy reach at least $10^{-11}$s). Hence, it is necessary to develop the metric up to the 2 PN/RM order. Our work is completely based on General Relativity (GR) and hence we don't consider any alternative theory in this paper. This means that our result would have a practical usefulness in telemetry only if the GR theory is the ''true'' theory of gravity or if alternative theory deviates from GR only by sufficiently small post-Newtonian numerical parameters in the metric. This is compatible with present available data.  \\
We start our work from the ''exponential parametrization'' such as in DSX \cite{DSX1991} or in the IAU2000 resolutions \cite{IAU2000}. Hence, our notations come from \cite{DSX1991}. Other works on the subject have already be done (\cite{Blanchet25PM},\cite{Anderson}) but they didn't write their metric in the convenient exponential parametrization and these works have been done in harmonic gauge from the start to the end. This is not our case since we fix the gauge only at the end when we want peculiar solutions in some peculiar gauges.\\
In section \ref{sec:term} we introduce a new terminology before recalling some well known results from the DSX paper \cite{DSX1991} in section \ref{sec:1}. Then we derive Einstein's equations in our metric's parametrization and finally we give formal solutions in any gauge that respect the exponential parametrization -- with a particular attention to the harmonic gauge.

\section{Terminology : definition of the PN/BM and PN/RM metrics}
\label{sec:term}

The PN approximation is based on the assumption of a weak gravitational
field and weak velocities (ie. of the order $\sqrt{GM/r}$ or less, $M$
being some caracteristic mass of the system) for both the sources and the
(test) body. It formally consists in looking for solutions under the form of
an expansion in powers of $1/c$. The usually so-called $n$PN order terms in the metric, leading to $%
c^{-2n}$ terms in the equation of motion of a body describing a bounded orbit, are terms
of orders $c^{-2n-2}$ in $g_{00}$, $c^{-2n-1}$ in $g_{0i}$ and $c^{-2n}$ in $%
g_{ij}$. In this paper, a metric developped this way will be refered as the nPN/BM metric (BM meaning "Bounded Motion" \ for test particles). It is particularly well-adapted for
studying bounded motions in systems made by non-relativistic massive bodies, as the
Solar System is. 

However, since we are interested in the propagation of light, we are lead to relax the hypothesis on the velocity of the test particle whose motion is considered. Of course, this doesn't change the full metric, but the terms to be
considered in the metric components are not the same as in the PN/BM
problem.\ Indeed, the terms leading to $c^{-2n}$ terms in the equation of
motion of a test particle moving with relativistic velocity  are terms of
order $c^{-2n}$ in $g_{\alpha \beta }$ (ie. in both $g_{00}$, $g_{0i}$
and $g_{ij}$). In this paper, a metric developped this way will be refered as the nPN/RM
metric (RM meaning "Relativistic Motion" \ for test particles). A PN/RM
metric is particularly well-suited for studying \textit{relativistic} motions of test bodies (for instance, the propagation of light) in systems made by non-relativistic bodies, as the Solar System is. 

\begin{figure}
\includegraphics[scale=0.25]{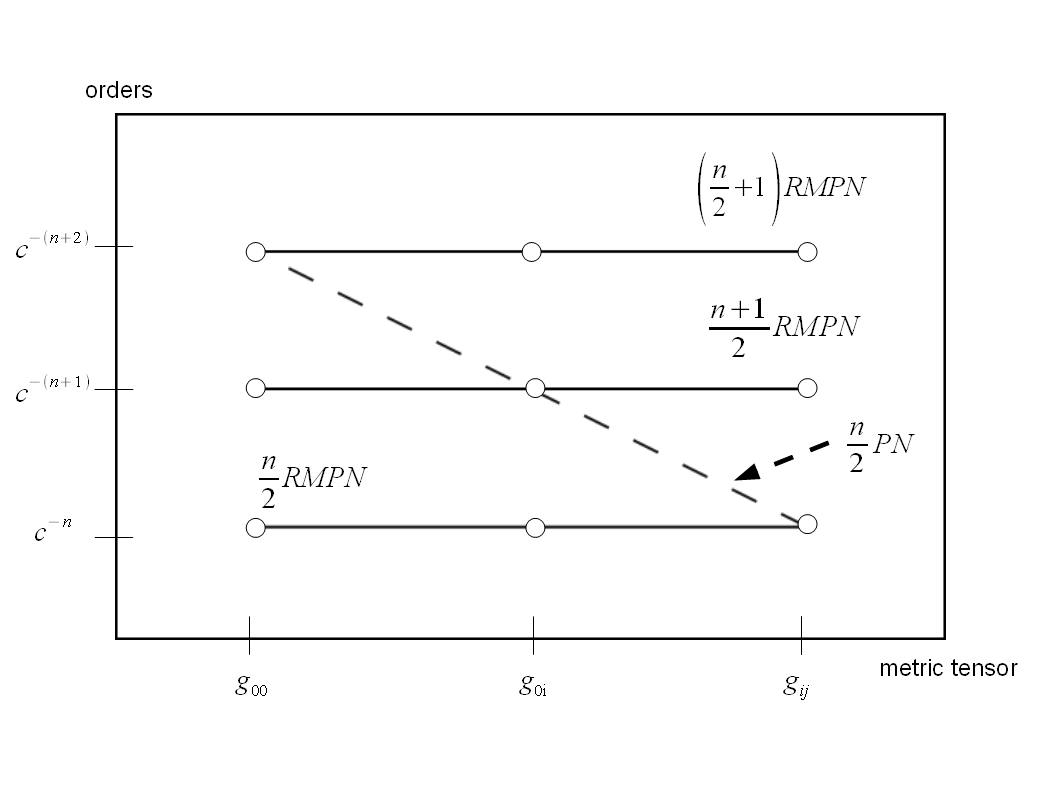}
\caption{General scheme of orders taken into account in PN/BM and PN/RM metrics.}
\label{fig:PMvsPN62}
\end{figure}

PN/BM and PN/RM orders are illustrated in figure \ref{fig:PMvsPN62}.

The present paper deals with the PN/RM problem since we are concerned in propagation of light.

\section{1,5 PN/BM metric}
\label{sec:1}

We recall here results from DSX\cite{DSX1991} on the 1,5PN/BM metric. First, the ''exponential parametrization'' with the spatial isotropy condition makes the metric writes

\begin{eqnarray}
g_{00}&=&-e^{-2w/c^2}\\
g_{0i}&=&-\frac{4}{c^3} w^i \\
g_{ij}&=&\gamma_{ij} e^{2w/c^2}.
\end{eqnarray}
With $\gamma_{ij}=\delta_{ij}+O(c^{-4})$. From which follows
\begin{eqnarray}
g^{00}&=&-e^{2w/c^2}+O(c^{-6})\\
g^{0i}&=&-\frac{4}{c^3} w_i+O(c^{-5})\\
g^{ij}&=&\delta^{ij} e^{-2w/c^2}+O(c^{-4}).
\end{eqnarray}
The Einstein field equation then reduces to four equations on $w$ and $w_i$
\begin{eqnarray}
\triangle w+\frac{3}{c^2}\partial^2_t w +\frac{4}{c^2} \partial_t \partial_j w_j = - 4 \pi G \sigma + O(c^{-4})\\
\label{eq:wi2PN}
\triangle w_i -\partial^2_{ij} w_j - \partial_t \partial_i w = - 4 \pi G \sigma^i + O(c^{-2}).
\end{eqnarray}
Where $\sigma=c^{-2} \left(T^{00}+T^{kk}\right)$, $\sigma^i = c^{-1} T^{0i}$ and $\triangle=\vec{\partial} \cdot \vec{\partial}$. Hence, to the 1,5PN/BM approximation correspond 4 equations of the GR equation and the gauge invariance keep 1 degree of freedom (indeed $\sigma^\mu$ is approximately divergence free). This gauge invariance is characterized by the arbitrary differentiable function $\lambda$ in DSX.
\begin{eqnarray}
w'=w-\frac{1}{c^2} \partial_t \lambda\\
w'_i=w_i+\frac{1}{4} \partial_i \lambda.
\end{eqnarray}
The IAU2000 resolution then corresponds to a peculiar choice of gauge in order to fix the last degree of freedom : the harmonic gauge. This condition writes, in the 1,5PN/BM approximation $\{g^{\alpha \beta}\Gamma^0_{\alpha \beta}=O(c^{-5}),g^{\alpha \beta}\Gamma^i_{\alpha \beta}=O(c^{-4})\}$. The four metric field equations then reduce to
\begin{eqnarray}
\label{eq:2PNw}
\Box_m w = - 4 \pi G \sigma + O(c^{-4})\\
\label{eq:2PNwi}
\triangle w_i = - 4 \pi G \sigma^i + O(c^{-2})
\end{eqnarray}
where $\Box_m= \triangle - \partial^2_t$.

\section{2 PN/RM metric}

At the 2PN/RM order, we get 6 new degrees of freedom in the metric. As we know, from the 10 degrees of freedom in the equation of GR, there are only 6 cleared by the 10 equations since 4 degrees of freedom correspond to a gauge invariance. Indeed, the metric is determined by the field equations modulo the knowing of the used coordinates. Since, there are four coordinates, there are 4 degrees of freedom not reduced by the 10 field equations. As seen in the previous section, to the 1,5PN/BM approximation correspond 4 metric field equations and the gauge invariance, characterized by $\lambda$, keep 1 degrees of freedom. Then, we expect from the 2PN/RM order 6 equations with a gauge invariant field (characterized by a 3-vector) which (the equations) reduce the 6 degrees of freedom to 3. Let us fix the covariant form of the metric

\begin{eqnarray}
g_{00}=-e^{-2w/c^2}\\
g_{0i}=-\frac{4}{c^3} w^i
\end{eqnarray}
\begin{multline}
g_{ij}=\delta_{ij}e^{2w/c^2} + \frac{4 \tau_{ij}}{c^4}+O(c^{-5})\\ = \delta_{ij} \left( 1+ \frac{2w}{c^2}+ \frac{2w^2}{c^4}\right) +   \frac{4 \tau_{ij}}{c^4}+O(c^{-5}).
\end{multline}

Which implie

\begin{eqnarray}
g^{00}=-e^{2w/c^2}+O(c^{-6})\\
g^{0i}=-\frac{4}{c^3} w_i+O(c^{-5})\\
g^{ij}=\delta^{ij} e^{-2w/c^2}-\frac{4 \tau_{ij}}{c^4}+O(c^{-5}).
\end{eqnarray}
Then we can derive the 2PN/RM field equations from the Einstein field equation $R^{\alpha \beta}=\frac{8 \pi G}{c^4} \left( T^{\alpha \beta} - \frac{1}{2} g^{\alpha \beta} g_{\rho \sigma} T^{\rho \sigma} \right)$.

\begin{multline}
\frac{2}{c^4} \Theta_{ij}\left( \tau_{kl} \right) + \Xi_{ij}\left(w,w_k \right) = \\ \frac{8 \pi G}{c^4} \left( \frac{1}{2} \eta^{ij} T^{00} + T^{ij} - \frac{1}{2} \eta^{ij} T^{kk} - \frac{2w}{c^2} \eta^{ij} T^{00} \right).
\end{multline}
Where
\begin{equation}
\Theta_{ij}\left( \tau_{kl} \right)=\partial^2_{ki} \tau_{k j}+\partial^2_{k j} \tau_{ik} - \triangle \tau_{ij} - \partial^2_{ij} \tau_{kk},
\end{equation}
\begin{multline}
\Xi_{ij}\left(w,w_k \right)= \\-\frac{\delta_{ij}}{c^2} \Box_m w + \frac{2}{c^3} \left( \partial^2_{0i} w_j + \partial^2_{0j} w_i  \right) - \frac{2}{c^4} \partial_i w \partial_j w + \frac{4 \delta_{ij}}{c^4} w \triangle w.	
\end{multline}

Then, considering $w$ and $w_k$ as source terms and identifying the relation $-\frac{c^4}{2}\left( \frac{4 \delta_{ij}}{c^4} w \triangle w \right)= \frac{8 \pi G}{c^2} \eta^{ij} w T^{00}+O(c^{-2})$ from (\ref{eq:2PNw}), we get

\begin{multline}
\Theta_{ij}\left( \tau_{kl} \right)=4 \pi G \sigma^{ij} \\- \left[ \left( \partial^2_{ti} w_j + \partial^2_{tj} w_i - \partial_i w \partial_j w   \right) + 2 \eta^{ij} \left(\partial^2_{t} w + \partial^2_{tk} w_k \right) \right].
\end{multline}

Where $\sigma^{ij}=T^{ij}-\eta^{ij}T^{kk}$. As expected, the operator $\Theta$ is gauge invariant through a relation on any 3-vector field. Indeed, we have $\Theta_{ij}\left( \tau_{kl} + \partial_k A_l + \partial_l A_k \right)=\Theta_{ij}\left( \tau_{kl} \right)$, whatever is the 3-vector field $A_k$. Hence, the 2PN/RM gauge invariance is characterized by both the arbitrary differentiable function $\lambda$ and the arbitrary differentiable 3-vector $A_k$.
\begin{eqnarray}
w'=w-\frac{1}{c^2} \partial_t \lambda\\
w'_i=w_i+\frac{1}{4} \partial_i \lambda\\
\tau'_{kl}=\tau_{kl} + \partial_k A_l + \partial_l A_k + \frac{1}{2} \delta_{kl} \partial_t \lambda.
\end{eqnarray}

\subsection{General formal solution}

We can show that it always exists a gauge transformation $\tau'_{kl}=\tau_{kl}+\partial_{k} A_l+\partial_{l} A_k$ that reduces the operator $\Theta$ to a Laplacian ($\Theta_{ij}\left( \tau'_{kl} \right)=-\triangle \tau'_{ij} $). To be placed on this gauge, $A_k$ must satisfy $\triangle A_j= \partial_k \tau_{kj} - \frac{1}{2} \partial_j \tau_{kk}$ which is clearly inversible since we know the laplacian's green function. Hence, there always exists a peculiar gauge where the solution is
\begin{multline}
\tau_{ij}= \triangle^{-1} \{ -4 \pi G \sigma^{ij} \\ + \left[ \left( \partial^2_{ti} w_j + \partial^2_{tj} w_i - \partial_i w \partial_j w   \right) + 2 \eta^{ij} \left(\partial^2_{t} w + \partial^2_{tk} w_k \right) \right]  \},
\end{multline}
where
\begin{equation}
\triangle^{-1} \{f(t,\vec{x}) \} = \int \frac{f(t,\vec{x}')}{|\vec{x}-\vec{x}'|} d^3 x'.
\end{equation}
Then, from this equation, we can get the solution $\tau_{kl}$ in any gauge, using any vector field $V_k$ with $\tau'_{kl}=\tau_{kl}+\partial_{k} V_l+\partial_{l} V_k$.

\subsection{The harmonic gauge}
The harmonic gauge condition stands $g^{\alpha \beta}\Gamma^\gamma_{\alpha \beta}=0$ (Or $\partial_\sigma \left( \sqrt{-g} g^{\sigma \gamma} \right)=0$ with $g$ the metric determinant). As seen in DSX \cite{DSX1991}, the harmonic 1,5PN/BM condition $g^{\alpha \beta}\Gamma^0_{\alpha \beta}=O(c^{-4})$ gives the usual harmonic constraint on the ''potentials'' ($\partial_t w + \partial_k w_k=O(c^{-2})$). While, the conditions $g^{\alpha \beta}\Gamma^i_{\alpha \beta}=O(c^{-4})$ give no more information due to the ''isotropy condition'' which makes spatial coordinates always harmonic modulo $O(c^{-4})$. At the 2PN/RM level, the harmonic constraint is $g^{\alpha \beta}\Gamma^\gamma_{\alpha \beta}=O(c^{-5})$. That gives the harmonic conditions on the ''potentials'' at the 2PN/RM level

\begin{eqnarray}
\label{eq:2PNharmonic}
\partial_t w + \partial_k w_k=O(c^{-2})\\
\partial_k \tau_{ik}-\frac{1}{2} \partial_i \tau{kk}+\partial_t w_i=O(c^{-1}).
\end{eqnarray}

Then $\Theta_{ij}\left( \tau_{kl} \right)$ writes:

\begin{multline}
\Theta_{ij}\left( \tau_{kl} \right) = - \{ \triangle \tau_{ij} + \partial_t \left(\partial_i w_j + \partial_j w_i \right) \}.
\end{multline}

Hence, considering $w$ and $w_i$ as source terms, the solution of the Einstein field equations in the harmonic gauge is :

\begin{equation}
\tau_{ij}= \triangle^{-1} \left\{ -4 \pi G \sigma^{ij}  - \partial_i w \partial_j w      \right\},
\end{equation}
where $w$ and $w_j$ satisfy the 1,5PN/BM harmonic condition (\ref{eq:2PNharmonic}).
\\\\
\textit{from non-harmonic to harmonic gauge}
\\\\
If $\tau_{ij}$ is not an harmonic solution, then the gauge transformation $\tau_{ij}=\epsilon_{ij}+\partial_i A_j + \partial_j A_i$ makes $\epsilon_{ij}$ an harmonic solution if $A_j$ satisfies

\begin{equation}
\triangle A_j = \partial_k \tau_{kj} - \frac{1}{2} \partial_j \tau_{kk} + \partial_t w_j
\end{equation}
where $w_j$ must be expressed in the harmonic gauge with the 1,5PN/BM field equations (\ref{eq:2PNwi}). Since it is inversible, this relation always allow us to put the metric in the harmonic gauge, whatever is the gauge from which we start.

\section{Conclusions}

In order to fit future inter-planetary telemetry data, we developed a GR metric up to the 2 PN/RM order with the DSX ''exponential parametrization''. We gave both general and harmonic formal solutions. A more detailed version of this result can be found in \cite{PRDMC}. An explicit solution of these equations and of the (non-massive-)Scalar-Tensor 2PN/RM field equations in the case of the solar system is in preparation.

\begin{acknowledgments}
Olivier Minazzoli wants to thank the Government of the Principality of Monaco for their financial support.
\end{acknowledgments}

\end{document}